\documentclass[12pt,letterpaper]{article}
\usepackage[top=1in,bottom=1in,left=1in,right=1in]{geometry}
\usepackage[utf8]{inputenc}
\usepackage[dvipsnames]{xcolor}
\usepackage{wrapfig}
\usepackage{amsmath}
\usepackage{amsfonts,amssymb}
\usepackage[square,numbers]{natbib}
\usepackage[pdftex]{graphicx}
\usepackage{wrapfig}
\usepackage{bold-extra}

\usepackage[colorlinks=true,linkcolor=black,citecolor=purple,urlcolor=purple]{hyperref}

\newcommand{\ue}[1]{\underline{\emph{#1}}}
\newcommand{\CII}{C~\textsc{ii}}
\newcommand{\OIII}{O~\textsc{iii}}
\newcommand{\NII}{N~\textsc{ii}}
\newcommand{\OI}{O~\textsc{i}}

\newcommand{\HII}{H~\textsc{ii}~}

\def\ben{\begin{enumerate}}
\def\een{\end{enumerate}}
\def\bi{\begin{itemize}}
\def\ei{\end{itemize}}
\def\be{\begin{equation}}
\def\ee{\end{equation}}
\def\bea{\begin{eqnarray}}
\def\eea{\end{eqnarray}}

\def\Lya{Ly$\alpha$~}
\def\Ha{H$\alpha$~}
\def\Hb{H$\beta$~}

\begin{document}
{\raggedright
\Large
Astro2020 Science White Paper \linebreak

{\bf Tomography of the Cosmic Dawn and Reionization Eras with Multiple Tracers} \linebreak
\normalsize

\noindent \textbf{Thematic Areas:} 
\hspace*{40pt} $\square$ Planetary Systems \hspace*{10pt} $\square$ Star and Planet Formation \hspace*{20pt}\linebreak
$\square$ Formation and Evolution of Compact Objects \hspace*{10pt} $\boxtimes$ Cosmology and Fundamental Physics \linebreak
  $\square$ Stars and Stellar Evolution \hspace*{1pt} $\square$ Resolved Stellar Populations and their Environments \linebreak
  $\boxtimes$ Galaxy Evolution   \hspace*{45pt} $\square$             Multi-Messenger Astronomy and Astrophysics \hspace*{65pt} \linebreak

\textbf{Principal Author:}

Name:	Tzu-Ching Chang
 \linebreak						
Institution:  Jet Propulsion Laboratory, California Institute of Technology
 \linebreak
Email: tzu-ching.chang@jpl.nasa.gov
 \linebreak
Phone:  626-298-5446
 \linebreak
 
{\textbf Co-authors:} 

Angus Beane (U Penn), Olivier Dor{\'e} (JPL), Adam Lidz (U Penn), Lluis Mas-Ribas (JPL), Guochao Sun (Caltech), Marcelo Alvarez (UC Berkeley), Ritoban Basu Thakur (Caltech), Philippe Berger (JPL), Matthieu Bethermin (LAM), Jamie Bock (Caltech), Charles M. Bradford (JPL), Patrick Breysse (CITA), Denis Burgarella (Aix-Marseille), Vassilis Charmandaris (Crete), Yun-Ting Cheng (Caltech), Kieran Cleary (Caltech), Asantha Cooray (UCI), Abigail Crites (Caltech),  Aaron Ewall-Wice (JPL), Xiaohui Fan (Arizona), Steve Finkelstein (UT Austin), Steve Furlanetto (UCLA), Jacqueline Hewitt (MIT), Jonathon Hunacek (Caltech), Phil Korngut (Caltech), Ely Kovetz (Ben-Gurion), Gregg Hallinan (Caltech), Caroline Heneka (SNS), Guilaine Lagache (LAM), Charles Lawrence (JPL), Joseph Lazio (JPL),  Adrian Liu (McGill), Dan Marrone (Arizona), Aaron Parsons (UC Berkeley), Anthony Readhead (Caltech), Jason Rhodes (JPL), Dominik Riechers (Cornell), Michael Seiffert (JPL), Gordon Stacey (Cornell), Eli Visbal (Flatiron), Hao-Yi Wu (OSU), Michael Zemcov (RIT), Zheng Zheng (Utah) 
\linebreak

{\textbf Endorsers:} 
James Aguirre (U Penn), Judd Bowman (ASU), Bryna Hazelton (UW), Kirit	Karkare (Chicago), Saul Kohn (Vanguard), Kiyoshi Masui (MIT), Jordan Mirocha (McGill), Laura Newburgh (Yale), Jonathan Pritchard (Imperial), John-David Smith (Toledo), Hy Trac (CMU)
\linebreak

{\textbf Description:} 
We advocate a large-scale, multi-tracer approach to study the reionization and cosmic dawn eras. We highlight the "line intensity mapping" technique to trace the multi-phase reionization topology on large scales, and measure reionization history in detail. We also advocate for Lya tomography mapping as an additional probe besides 21cm of the cosmic dawn era.

}

\pagebreak

\pagenumbering{gobble}
\newpage
\pagenumbering{arabic}

\vspace{-0.5cm}
\section{Probing the Epochs of Reionization and Cosmic Dawn}

The development of a comprehensive understanding of the physics that led to the formation of the first stars and galaxies -- the Cosmic Dawn and the Epoch of Reionization (EoR) -- is a challenging but fundamental goal of astrophysics and cosmology. Although tremendous progress has been made in the past two decades, many key questions remain. For example, how and when did the first UV-bright stars that reionized the universe form from pristine primordial gas?  What is the history of stellar, dust and metal build-up during reionization?  What is the contribution of quasars and AGN to the reionization history of the universe? \emph{Answering these questions during the EoR requires multi-phase tracers of interstellar and intergalactic gas across a broad range of spatial scales. Multi-emission line surveys of the EoR will provide the necessary probes.} 

\vspace{-0.5cm}
\paragraph{}
One of the major challenges for reionization studies is the enormous dynamic range in spatial scale involved; reionization directly couples the very small-scale astrophysical formation of stars to the cosmological length scales of structure formation. This involves complex astrophysical processes such as the propagation of ionizing photons through interstellar, circumgalactic, and intergalactic gas. 
This presents challenges for both numerical simulations and especially observations. Current major observational efforts include measurements of individual high-redshift sources
 with HST, ALMA and other ground-based observatories on small patches of sky, whereas the measurement of the optical depth to electron scattering from Planck gives an integral constraint over the entire reionization history. This information is sparse and not matched in scale to the reionization phenomena, driven by the growth and percolation of ionized regions (bubbles) that are of Mpc or 10s of Mpc in scale.  On-going and upcoming 21~cm reionization experiments in the 2020s such as HERA and SKA1-LOW will statistically map the neutral intergalactic medium as traced by the 21~cm transition in three dimensions, with the hope of directly imaging the neutral IGM towards the 2030s. The 21~cm surveys will cover tens to thousands of square degrees, and can therefore capture representative samples of the ionized bubbles. However, these surveys track just the neutral hydrogen gas and provide only indirect constraints on the formation and evolution of the ionizing sources themselves.
 
\vspace{-0.5cm}
\paragraph{}
Here we highlight how the \emph{Line Intensity Mapping} (LIM) approach may provide an important key to the reionization puzzle. Much like the approach of 21cm EoR surveys but more generally, LIM measures the collective emission of a transition line on Mpc scales and larger. Since the line in use traces directly the neutral, partially ionized, or ionized intergalactic medium, LIM provides crucial environmental and statistical measurements of the IGM and large-scale structure that complement galaxy detections.  It also allows direct inferences of the cosmic history of metal and dust build-up and the history of reionization. By combining multiple such line measurements, originating from different phases of the ISM and the IGM, one can determine the large-scale average physical conditions at early times and reveal the physical processes driving reionization. We also advocate for additional probes of the cosmic dawn era, when early stars had yet to significantly enrich their surroundings with metals. Besides the 21~cm transition from neutral hydrogen, \Lya tomography mapping during the epoch of Wouthuysen-Field (WF) coupling may potentially be detectable as a faint glow from cosmic dawn.

\vspace{-0.5cm}
\section{The Topology and History of Reionization}

The cosmic reionization of hydrogen remains a poorly understood frontier of modern cosmology.
Sometime between 200 and 800 Myrs after the Big Bang ($z\sim$ 6-20), the first luminous objects formed in dark matter halos and eventually produced enough Lyman continuum photons to reionize the surrounding hydrogen gas. The EoR marked the end of the dark ages, and is the first chapter in the history of galaxies and heavy elements \cite{BL_2001}. Observations of the Gunn-Peterson trough \cite{GP_1965} in high-redshift quasar spectra suggest that reionization was completed by $z \simeq 6$ \cite{Becker_2001, Fan_2006}. Current constraints from large-angle cosmic microwave background (CMB) polarization measurements indicate an optical depth to reionization of $\tau=$ 0.054 $\pm$ 0.007, corresponding to an instantaneous reionization redshift of 7.61 $\pm$ 0.75 \cite{Planck_VI_2018}.  Although these observations provide some constraints on the overall timing of reionization, it is likely an extended and inhomogeneous process and so a great deal remains to be understood. We anticipate that the coming decade is ripe for fundamental breakthroughs.

\vspace{-0.5cm}
\paragraph{Studying the Reionization Epoch by Resolving its Sources.} Direct observations of galaxies and quasars beyond $z \simeq 6$
provide tight constraints on crucial stages of reionization. 
Observing these sources has been a major effort of the HST. For example, broad-band searches for $z \simeq 6$-10 dropout galaxies (or Lyman
Break Galaxies (LBGs)) have led to the identification of $> 1000$ candidates. 
In the near future, JWST will directly observe galaxies in the EoR. However, the JWST cosmological fields will be limited to a handful of deep fields, with a total area of several hundred square arcmin and hence subject to sample variance. Soon after, due to its large and deep survey capabilities, WFIRST will detect several million 
LBGs at a redshift $z>6$ over the $\simeq$ 2,000 sq. degree High Latitude Survey (HLS) \cite{Waters_2016}. Deep GO grism program will be able to access thousands of Ly$\alpha$ emitters (LAE), albeit likely limited to small fields. While WFIRST will perform wide area surveys, its lack of capability beyond 2 $\mu$m limits measurements of stellar mass (in the rest-frame optical) and spectroscopic detection of H$\alpha$ at $z > 2$, and the selection of $z > 6$ galaxies will be limited to photometric data. 
Although JWST and WFIRST will provide important information about early galaxy populations, they will suffer from incompleteness at the faint end of the luminosity function and should also be supplemented by wide-field spectroscopic surveys. Fortunately, other complementary approaches have emerged.

\vspace{-0.5cm}
\paragraph{Studying Reionization by Mapping the Collective Light on Large Scales.} Many on-going experiments
now aim at mapping the EoR using LIM in multiple lines
such as the radio hyperfine 21 cm line (e.g, LOFAR, PAPER, MWA, HERA, SKA1-LOW),
CO (e.g., COMAP, AIM-CO), [\CII] (e.g. TIME, Concerto, CCAT-prime), and H$\alpha$ and \Lya (SPHEREx; see \cite{Kovetz_2017} for references). Without resolving individual luminous sources, line intensity mapping (LIM) measures all of the light emissions of a particular spectral line associated with cosmic structures on large scales. This IGM- and larger-scale information provides the environmental context and connects directly to individual source measurements. LIM has emerged as a promising approach to efficiently survey the three dimensional universe, probing the large-scale structure (LSS), EoR and the Cosmic Dawn era \cite{Chang_2008, WL_2008, VL_2010}. 
\emph{LIM measures the cumulative emission of light from all galaxies and can be used to (1) infer the mean luminosity density and redshift evolution in lines such as CO, [\CII], [\OIII], and \Ha, (2) place constraints on the early star-formation history (inferred through the luminosity density in various lines), and (3) as a tracer of the reionization history (via its connection with the star-formation history.)}

\begin{wrapfigure}{r}{0.42\textwidth}
\vspace{-1cm}
\begin{center}
\includegraphics*[width=0.42\textwidth]{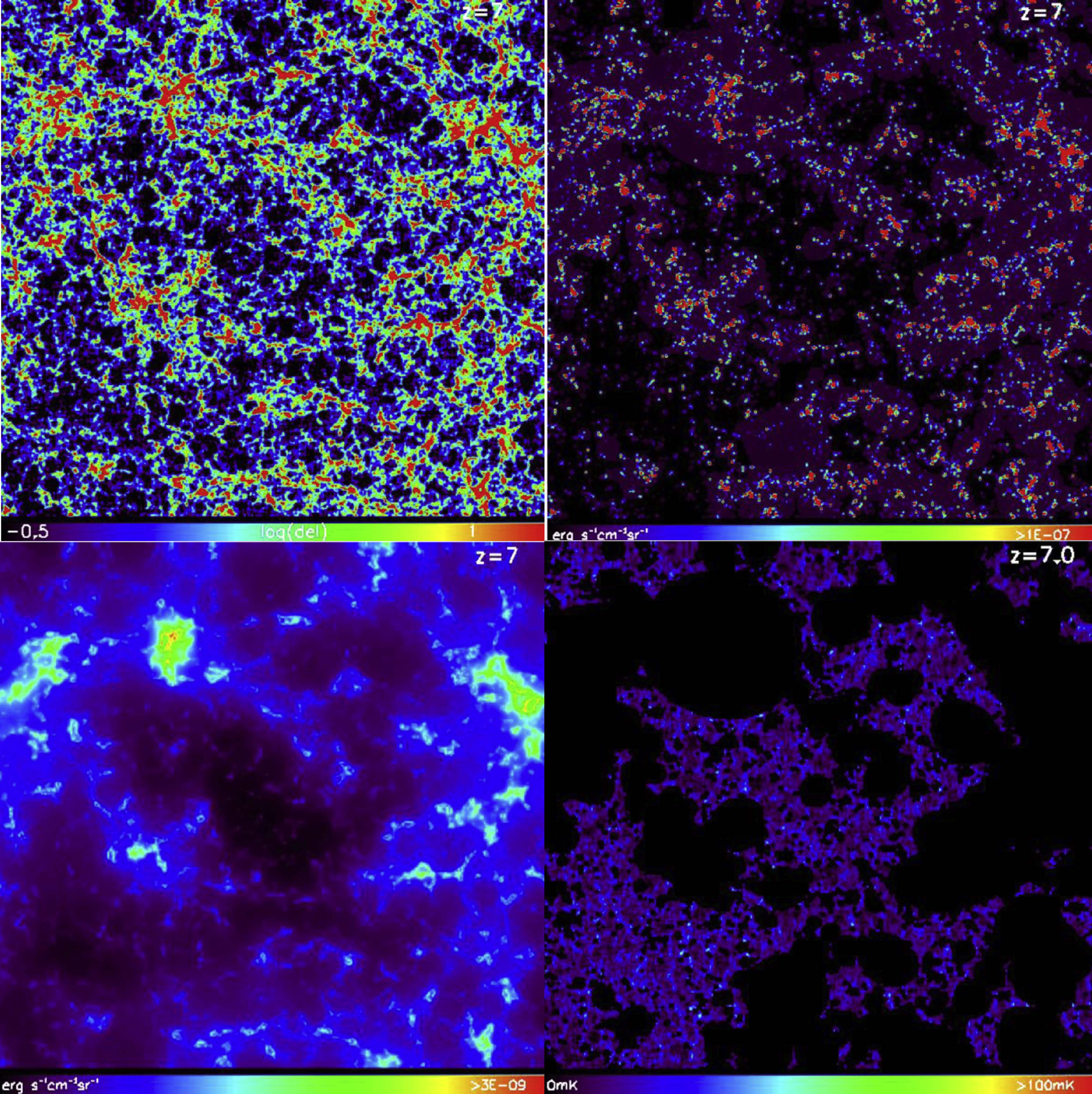}
\caption{\small{Different phases of reionization at $z=7$ from simulations \cite{Heneka17}, where the box size in each panel is 200 Mpc. Clock-wise from top left is the dark matter density field, the ionizing sources as traced by \Ha, the neutral IGM traced by 21~cm, and the partially ionized IGM traced by \Lya scattering.
    LIM provides information on large spatial-scales, comparable to those probed by 21 cm surveys.
    Combining large-scale tomographic intensity maps in three lines (\Ha, 21~cm and \Lya) therefore provides a comprehensive view of reionization and reveals the topology and history of reionization.}} 
\label{fig:reionization}
\vspace{-1cm}
\end{center}
\end{wrapfigure}

\vspace{-0.5cm}
\paragraph{Multi-LIM Approach.} Studying multiple lines with LIM  simultaneously offers a richer, more nuanced and more robust view of the processes at play during the EoR (Fig.~\ref{fig:reionization}). 
Specifically, we hope to access bright emission lines in the rest-frame UV/optical window, as well as near-infrared and radio lines.
While the insights gained by joint analyses of these lines have been well studied on individual source scales, we discuss examples below to motivate their studies on large scales during the EoR, in particular in the intensity mapping regime as {\it luminosity-weighted} physical quantities.
In addition to science gain, we expect the joint analysis of multiple lines will greatly aid the confirmation of measured signals and help mitigate modeling and instrumental systematic effects, as each line arises from different physical processes, can be contaminated by different astrophysical foreground emissions, and will be observed through different instruments.  

\vspace{-0.5cm}
\paragraph{Key Emission Lines.} \ue{Optical and UV lines.} The ISM in the neighborhood of young, massive stars is ionized by the Lyman
continuum photons produced by these stars, thereby giving rise to \HII regions. The recombination of this ionized gas produces hydrogen emission lines (e.g., Ly$\alpha$, H$\alpha$, H$\beta$), which can be used as a Star Formation Rate (SFR) diagnostic since the line flux is proportional to the Lyman
continuum flux. Since the H$\alpha$ luminosity is a measure of the formation rate of massive stars, the derived SFR is sensitive to the
assumed stellar initial mass function (IMF). Dust
extinction may cause systematic errors due to non-negligible dust
absorption at the \Ha wavelength, however, if the \Hb line is also measured, one may use the ratio to
correct for dust extinction. 
\ue{Far-infrared lines.}  These optical/UV lines are nicely complemented by ISM emission lines 
from fine-structure transitions of ionized carbon [\CII] (157.7 $\mu$m), ionized nitrogen [\NII] (121.9 $\mu$m and
205.2 $\mu$m) and neutral oxygen [OI] (145.5 $\mu$m) at redshift z $>$ 4, as well as the 88 $\mu$m [\OIII] line now seen at high-redshift. These lines are important coolants of both the neutral and the ionized medium, and jointly probe different phases of the ISM, which is key to understand the interplay between energetic sources, and the gas and dust at high redshift.
\ue{Radio lines.} Stars form in clouds of cold molecular hydrogen, but because H$_2$ is invisible in the cold ISM, its distribution and motion is inferred from observations of minor constituents of the clouds, such as CO and dust. Mapping the CO thus gives us a direct survey of the molecular regions where star forms. 
Finally, the 21 cm hyperfine transition of neutral hydrogen has become one of the most exciting probes of reionization. Detecting intensity fluctuations in the redshifted 21 cm signal arising from variations in the neutral fraction allows us to probe the spatial inhomogeneities of the reionization process. 
\emph{Taken together, observations of these lines constrain the properties of the intergalactic medium (IGM) and the cumulative impact of light from all galaxies, integrated over the bright and faint end of the luminosity function. In combination with direct
observations of the sources they provide a powerful tool for learning about the first stars and galaxies.} 

\vspace{-0.5cm}
\paragraph{Multi-phase IGM, Ionized Bubble Size, and Reionization History.}\ue{H$\alpha$, Ly$\alpha$, and} \ue{21~cm.} Intensity maps of H$\alpha$, Ly$\alpha$, and 21cm radiation jointly provide a comprehensive view of the topology and history of reionization (Fig.~\ref{fig:reionization}). \Lya photons can be produced through recombinations, and appear more extended than the stellar continuum when the ionizing radiation escapes the stellar sites and ionizes distant neutral gas.  \Lya can additionally be created by collisional excitation of neutral gas falling into the galactic potential wells, heated by gravitational energy and cooled through radiative \Lya emission when the gas metallicity is low. Furthermore, \Lya radiation diffuses spatially and in frequency as it scatters resonantly through surrounding neutral gas. 
\Lya thus traces the ionized and partially ionized IGM and appears more extended than H$\alpha$ emission, which is spatially associated with the ionizing sources. The 21~cm emission, on the other hand, traces the neutral IGM. Through cross-correlation studies, this multi-phase view of the reionization process can trace the detailed ionization history and the evolution of ionized bubble sizes \cite{Heneka17}.

\ue{[\CII] and 21~cm.} Similarly, as one of the strongest cooling lines of the ISM, the [\CII] 158 $\mu$m line has been shown to be a reliable tracer of star formation \cite{DeLooze_2014}. Multiple cross-correlations of [\CII] with for example [N II] and [O I] make it possible to extract the mean intensity of each individual line in a way minimally affected by uncorrelated foregrounds, through which physical properties of the ISM, such as the mean density of particles and gas temperature, may be constrained \cite{Serra_2016}. Measurements of the cross correlation of [\CII], \Lya, and 21~cm signals are useful for both overcoming low-redshift foregrounds of 21~cm data, and, more importantly, for constraining the average size of ionized bubbles by tracing the scale at which the cross correlation changes sign \cite{Lidz:2008ry,Lidz11,Gong_2012,Chang_2015,Dumitru_2018}. Multi-line cross-correlations have also been considered to robustly measure the fluctuations in the 21~cm signal \cite{2018arXiv181110609B}. Moreover, high-order statistics such as cross-bispectrum of [\CII] and 21~cm fields promise to help in reliably extracting 21~cm bias factors \cite{BL_2018}.  Recently, Fujimoto et al. \cite{Fujimoto_2019} reported a detection of extended [\CII] emission in halos at $z\sim 5-7$. If the spatial extent of \Lya overlaps with that of [\CII] 158\,$\mu$m and/or 88\,$\mu$m [\OIII], this will indicate that cooling is present and therefore one can distinguish the origin of \Lya from this process. 
\vspace{-0.5cm}
\section{Faint Glows from the Cosmic Dawn}

\vspace{-0.35cm}
\paragraph{} The Cosmic Dawn era here loosely refers to the early epoch of the formation of the first luminous objects at $z \sim 15$ before reionization. The ignition of the first collapsed objects lit up the universe in an otherwise pristine environment of hydrogen and helium. The high-energy photons escaped to heat up the surrounding medium on large-scales, while ionization and recombination also took place. The imprint of these early formation activities can be seen as a spectral distortion of the CMB at low frequencies, as indicated by the global 21~cm brightness temperature decrement against the CMB (reportedly detected by EDGES at $z\sim17$ \cite{Bowman_2018}). Independent of the EDGES result, prior to the redshift associated to the anticipated absorption signature, the 21~cm spin temperature must be coupled to the gas kinetic temperature;
this WF coupling happened when the UV photons from the first sources redshifted into the \Lya resonance: 
atoms may swap hyperfine states as they absorb and subsequently re-emit \Lya photons. In net, this couples the spin temperature of the 21~cm transition to the gas temperature. For the coupling to be effective, some minimum level of UV photons redshifting into the \Lya resonance is required.
The minimum level of \Lya photon production at early times
is on the order of one \Lya photon for every ten hydrogen atoms \cite{Chen:2003gc}, 
with the precise number depending on the coupling redshift and whether the incident high redshift radio background is enhanced above the CMB \cite{Ewall-Wice:2018bzf}.
Additionally, \Lya photons can be produced in the recombination cascades that take place within \HII regions in the early galaxies themselves, as well as those from recombinations within the IGM. These recombination produced \Lya photons may in fact dominate the overall average specific intensity at Ly$\alpha$ frequencies, even though they only appear close to sources and so are not useful for coupling the spin temperature to the gas temperature throughout most of the IGM. 

\vspace{-0.5cm}
\paragraph{}There are multiple on-going and planned efforts to measure the global, mean 21~cm brightness temperature against the CMB (e.g., EDGES, SARAS, PRIZM, LEDA) that will continue into the 2020s, as well as 21~cm fluctuations during the cosmic dawn (HERA, LWA, SKA1-LOW). The 21~cm field has been vibrant and promising, although the measurements are challenging in large part due to the presence of astrophysical foregrounds that are several orders of magnitude more prominent. \Lya photons may provide a promising, additional probe of cosmic dawn, revealing crucial clues to the early structure formation and possibly dark matter-baryon interactions, although this faint signal may be challenging to detect. Other possible, faint LIM probes of this epoch include the HeII 1640 A line \cite{Visbal15} and emission from molecular hydrogen \cite{Gong13}.

\vspace{-0.5cm}
\section{Future prospects}

\vspace{-0.35cm}
\paragraph{} The LIM program can be ideally facilitated by high-surface-brightness sensitivity, high-survey-speed instruments with wideband spectroscopic capability, along with advances in theoretical and numerical modeling and in particular data analysis techniques.  In the radio, current and on-going compact, large-collecting area instruments are already optimized for 21~cm tomography studies. In the mm, sub-mm, and far-infrared, LIM capability is poised for a revolution thanks to gains in detector sensitivity and format. On-chip spectrometers are nearing fruition for the ground-based millimeter bands; they will enable large spectrometer arrays accessing [\CII], [\NII] and [\OIII] in the EoR. Such technological advancements have crossovers with the active detector development program being pursued by the CMB community, and synergies between the CMB and LIM groups will enable rapid convergence for necessary instrumentation as well as data analysis techniques.  In the THz/far-IR, background-limited spectrographs on cryogenic space telescopes such as the Origins Space Telescope (or smaller variants) can enable measurements of [\OI] emission at $z=5-8$ and potentially H2 out to $z=15$.  In the near-IR, the SPHEREx mission presents the first major step towards realizing this goal, measuring \Ha at $z=0.5 -3$ and potentially the \Lya fluctuations during reionization.  Finally the Probe study concept Cosmic Dawn Intensity Mapper (CDIM) with enhanced sensitivity could potentially measure \Ha at $z=0.5-8$ and peek into the very high redshift universe via tomographic \Lya mapping. 

\newpage

\section*{Acknowledgement}
Part of this research was carried out at the Jet Propulsion Laboratory, California Institute of Technology, under a contract with the National Aeronautics and Space Administration. © 2019. All rights reserved.

\bibliography{refs}

\begin{thebibliography}{26}
\expandafter\ifx\csname natexlab\endcsname\relax\def\natexlab#1{#1}\fi
\expandafter\ifx\csname bibnamefont\endcsname\relax
  \def\bibnamefont#1{#1}\fi
\expandafter\ifx\csname bibfnamefont\endcsname\relax
  \def\bibfnamefont#1{#1}\fi
\expandafter\ifx\csname citenamefont\endcsname\relax
  \def\citenamefont#1{#1}\fi
\expandafter\ifx\csname url\endcsname\relax
  \def\url#1{\texttt{#1}}\fi
\expandafter\ifx\csname urlprefix\endcsname\relax\def\urlprefix{URL }\fi
\providecommand{\bibinfo}[2]{#2}
\providecommand{\eprint}[2][]{\url{#2}}

\bibitem[{\citenamefont{{Barkana} and {Loeb}}(2001)}]{BL_2001}
\bibinfo{author}{\bibfnamefont{R.}~\bibnamefont{{Barkana}}} \bibnamefont{and}
  \bibinfo{author}{\bibfnamefont{A.}~\bibnamefont{{Loeb}}},
  \bibinfo{journal}{\physrep} \textbf{\bibinfo{volume}{349}},
  \bibinfo{pages}{125} (\bibinfo{year}{2001}), \eprint{astro-ph/0010468}.

\bibitem[{\citenamefont{{Gunn} and {Peterson}}(1965)}]{GP_1965}
\bibinfo{author}{\bibfnamefont{J.~E.} \bibnamefont{{Gunn}}} \bibnamefont{and}
  \bibinfo{author}{\bibfnamefont{B.~A.} \bibnamefont{{Peterson}}},
  \bibinfo{journal}{\apj} \textbf{\bibinfo{volume}{142}}, \bibinfo{pages}{1633}
  (\bibinfo{year}{1965}).

\bibitem[{\citenamefont{{Becker} et~al.}(2001)\citenamefont{{Becker}, {Fan},
  {White}, {Strauss}, {Narayanan}, {Lupton}, {Gunn}, {Annis}, {Bahcall},
  {Brinkmann} et~al.}}]{Becker_2001}
\bibinfo{author}{\bibfnamefont{R.~H.} \bibnamefont{{Becker}}},
  \bibinfo{author}{\bibfnamefont{X.}~\bibnamefont{{Fan}}},
  \bibinfo{author}{\bibfnamefont{R.~L.} \bibnamefont{{White}}},
  \bibinfo{author}{\bibfnamefont{M.~A.} \bibnamefont{{Strauss}}},
  \bibinfo{author}{\bibfnamefont{V.~K.} \bibnamefont{{Narayanan}}},
  \bibinfo{author}{\bibfnamefont{R.~H.} \bibnamefont{{Lupton}}},
  \bibinfo{author}{\bibfnamefont{J.~E.} \bibnamefont{{Gunn}}},
  \bibinfo{author}{\bibfnamefont{J.}~\bibnamefont{{Annis}}},
  \bibinfo{author}{\bibfnamefont{N.~A.} \bibnamefont{{Bahcall}}},
  \bibinfo{author}{\bibfnamefont{J.}~\bibnamefont{{Brinkmann}}},
  \bibnamefont{et~al.}, \bibinfo{journal}{\aj} \textbf{\bibinfo{volume}{122}},
  \bibinfo{pages}{2850} (\bibinfo{year}{2001}), \eprint{astro-ph/0108097}.

\bibitem[{\citenamefont{{Fan} et~al.}(2006)\citenamefont{{Fan}, {Strauss},
  {Becker}, {White}, {Gunn}, {Knapp}, {Richards}, {Schneider}, {Brinkmann}, and
  {Fukugita}}}]{Fan_2006}
\bibinfo{author}{\bibfnamefont{X.}~\bibnamefont{{Fan}}},
  \bibinfo{author}{\bibfnamefont{M.~A.} \bibnamefont{{Strauss}}},
  \bibinfo{author}{\bibfnamefont{R.~H.} \bibnamefont{{Becker}}},
  \bibinfo{author}{\bibfnamefont{R.~L.} \bibnamefont{{White}}},
  \bibinfo{author}{\bibfnamefont{J.~E.} \bibnamefont{{Gunn}}},
  \bibinfo{author}{\bibfnamefont{G.~R.} \bibnamefont{{Knapp}}},
  \bibinfo{author}{\bibfnamefont{G.~T.} \bibnamefont{{Richards}}},
  \bibinfo{author}{\bibfnamefont{D.~P.} \bibnamefont{{Schneider}}},
  \bibinfo{author}{\bibfnamefont{J.}~\bibnamefont{{Brinkmann}}},
  \bibnamefont{and}
  \bibinfo{author}{\bibfnamefont{M.}~\bibnamefont{{Fukugita}}},
  \bibinfo{journal}{\aj} \textbf{\bibinfo{volume}{132}}, \bibinfo{pages}{117}
  (\bibinfo{year}{2006}), \eprint{astro-ph/0512082}.

\bibitem[{\citenamefont{{Planck Collaboration}
  et~al.}(2018)\citenamefont{{Planck Collaboration}, {Aghanim}, {Akrami},
  {Ashdown}, {Aumont}, {Baccigalupi}, {Ballardini}, {Banday}, {Barreiro},
  {Bartolo} et~al.}}]{Planck_VI_2018}
\bibinfo{author}{\bibnamefont{{Planck Collaboration}}},
  \bibinfo{author}{\bibfnamefont{N.}~\bibnamefont{{Aghanim}}},
  \bibinfo{author}{\bibfnamefont{Y.}~\bibnamefont{{Akrami}}},
  \bibinfo{author}{\bibfnamefont{M.}~\bibnamefont{{Ashdown}}},
  \bibinfo{author}{\bibfnamefont{J.}~\bibnamefont{{Aumont}}},
  \bibinfo{author}{\bibfnamefont{C.}~\bibnamefont{{Baccigalupi}}},
  \bibinfo{author}{\bibfnamefont{M.}~\bibnamefont{{Ballardini}}},
  \bibinfo{author}{\bibfnamefont{A.~J.} \bibnamefont{{Banday}}},
  \bibinfo{author}{\bibfnamefont{R.~B.} \bibnamefont{{Barreiro}}},
  \bibinfo{author}{\bibfnamefont{N.}~\bibnamefont{{Bartolo}}},
  \bibnamefont{et~al.}, \bibinfo{journal}{arXiv e-prints}
  \bibinfo{eid}{arXiv:1807.06209} (\bibinfo{year}{2018}), \eprint{1807.06209}.

\bibitem[{\citenamefont{{Waters} et~al.}(2016)\citenamefont{{Waters}, {Di
  Matteo}, {Feng}, {Wilkins}, and {Croft}}}]{Waters_2016}
\bibinfo{author}{\bibfnamefont{D.}~\bibnamefont{{Waters}}},
  \bibinfo{author}{\bibfnamefont{T.}~\bibnamefont{{Di Matteo}}},
  \bibinfo{author}{\bibfnamefont{Y.}~\bibnamefont{{Feng}}},
  \bibinfo{author}{\bibfnamefont{S.~M.} \bibnamefont{{Wilkins}}},
  \bibnamefont{and} \bibinfo{author}{\bibfnamefont{R.~A.~C.}
  \bibnamefont{{Croft}}}, \bibinfo{journal}{\mnras}
  \textbf{\bibinfo{volume}{463}}, \bibinfo{pages}{3520} (\bibinfo{year}{2016}),
  \eprint{1605.05670}.

\bibitem[{\citenamefont{{Kovetz} et~al.}(2017)\citenamefont{{Kovetz}, {Viero},
  {Lidz}, {Newburgh}, {Rahman}, {Switzer}, {Kamionkowski}, {Aguirre},
  {Alvarez}, {Bock} et~al.}}]{Kovetz_2017}
\bibinfo{author}{\bibfnamefont{E.~D.} \bibnamefont{{Kovetz}}},
  \bibinfo{author}{\bibfnamefont{M.~P.} \bibnamefont{{Viero}}},
  \bibinfo{author}{\bibfnamefont{A.}~\bibnamefont{{Lidz}}},
  \bibinfo{author}{\bibfnamefont{L.}~\bibnamefont{{Newburgh}}},
  \bibinfo{author}{\bibfnamefont{M.}~\bibnamefont{{Rahman}}},
  \bibinfo{author}{\bibfnamefont{E.}~\bibnamefont{{Switzer}}},
  \bibinfo{author}{\bibfnamefont{M.}~\bibnamefont{{Kamionkowski}}},
  \bibinfo{author}{\bibfnamefont{J.}~\bibnamefont{{Aguirre}}},
  \bibinfo{author}{\bibfnamefont{M.}~\bibnamefont{{Alvarez}}},
  \bibinfo{author}{\bibfnamefont{J.}~\bibnamefont{{Bock}}},
  \bibnamefont{et~al.}, \bibinfo{journal}{arXiv e-prints}
  (\bibinfo{year}{2017}), \eprint{1709.09066}.

\bibitem[{\citenamefont{{Chang} et~al.}(2008)\citenamefont{{Chang}, {Pen},
  {Peterson}, and {McDonald}}}]{Chang_2008}
\bibinfo{author}{\bibfnamefont{T.-C.} \bibnamefont{{Chang}}},
  \bibinfo{author}{\bibfnamefont{U.-L.} \bibnamefont{{Pen}}},
  \bibinfo{author}{\bibfnamefont{J.~B.} \bibnamefont{{Peterson}}},
  \bibnamefont{and}
  \bibinfo{author}{\bibfnamefont{P.}~\bibnamefont{{McDonald}}},
  \bibinfo{journal}{Physical Review Letters} \textbf{\bibinfo{volume}{100}},
  \bibinfo{eid}{091303} (\bibinfo{year}{2008}), \eprint{0709.3672}.

\bibitem[{\citenamefont{{Wyithe} et~al.}(2008)\citenamefont{{Wyithe}, {Loeb},
  and {Geil}}}]{WL_2008}
\bibinfo{author}{\bibfnamefont{J.~S.~B.} \bibnamefont{{Wyithe}}},
  \bibinfo{author}{\bibfnamefont{A.}~\bibnamefont{{Loeb}}}, \bibnamefont{and}
  \bibinfo{author}{\bibfnamefont{P.~M.} \bibnamefont{{Geil}}},
  \bibinfo{journal}{\mnras} \textbf{\bibinfo{volume}{383}},
  \bibinfo{pages}{1195} (\bibinfo{year}{2008}), \eprint{0709.2955}.

\bibitem[{\citenamefont{{Visbal} and {Loeb}}(2010)}]{VL_2010}
\bibinfo{author}{\bibfnamefont{E.}~\bibnamefont{{Visbal}}} \bibnamefont{and}
  \bibinfo{author}{\bibfnamefont{A.}~\bibnamefont{{Loeb}}},
  \bibinfo{journal}{\jcap} \textbf{\bibinfo{volume}{11}}, \bibinfo{eid}{016}
  (\bibinfo{year}{2010}), \eprint{1008.3178}.

\bibitem[{\citenamefont{{Heneka} et~al.}(2017)\citenamefont{{Heneka}, {Cooray},
  and {Feng}}}]{Heneka17}
\bibinfo{author}{\bibfnamefont{C.}~\bibnamefont{{Heneka}}},
  \bibinfo{author}{\bibfnamefont{A.}~\bibnamefont{{Cooray}}}, \bibnamefont{and}
  \bibinfo{author}{\bibfnamefont{C.}~\bibnamefont{{Feng}}},
  \bibinfo{journal}{\apj} \textbf{\bibinfo{volume}{848}}, \bibinfo{eid}{52}
  (\bibinfo{year}{2017}), \eprint{1611.09682}.

\bibitem[{\citenamefont{{De Looze} et~al.}(2014)\citenamefont{{De Looze},
  {Cormier}, {Lebouteiller}, {Madden}, {Baes}, {Bendo}, {Boquien}, {Boselli},
  {Clements}, {Cortese} et~al.}}]{DeLooze_2014}
\bibinfo{author}{\bibfnamefont{I.}~\bibnamefont{{De Looze}}},
  \bibinfo{author}{\bibfnamefont{D.}~\bibnamefont{{Cormier}}},
  \bibinfo{author}{\bibfnamefont{V.}~\bibnamefont{{Lebouteiller}}},
  \bibinfo{author}{\bibfnamefont{S.}~\bibnamefont{{Madden}}},
  \bibinfo{author}{\bibfnamefont{M.}~\bibnamefont{{Baes}}},
  \bibinfo{author}{\bibfnamefont{G.~J.} \bibnamefont{{Bendo}}},
  \bibinfo{author}{\bibfnamefont{M.}~\bibnamefont{{Boquien}}},
  \bibinfo{author}{\bibfnamefont{A.}~\bibnamefont{{Boselli}}},
  \bibinfo{author}{\bibfnamefont{D.~L.} \bibnamefont{{Clements}}},
  \bibinfo{author}{\bibfnamefont{L.}~\bibnamefont{{Cortese}}},
  \bibnamefont{et~al.}, \bibinfo{journal}{\aap} \textbf{\bibinfo{volume}{568}},
  \bibinfo{eid}{A62} (\bibinfo{year}{2014}), \eprint{1402.4075}.

\bibitem[{\citenamefont{{Serra} et~al.}(2016)\citenamefont{{Serra}, {Dor{\'e}},
  and {Lagache}}}]{Serra_2016}
\bibinfo{author}{\bibfnamefont{P.}~\bibnamefont{{Serra}}},
  \bibinfo{author}{\bibfnamefont{O.}~\bibnamefont{{Dor{\'e}}}},
  \bibnamefont{and}
  \bibinfo{author}{\bibfnamefont{G.}~\bibnamefont{{Lagache}}},
  \bibinfo{journal}{\apj} \textbf{\bibinfo{volume}{833}}, \bibinfo{eid}{153}
  (\bibinfo{year}{2016}), \eprint{1608.00585}.

\bibitem[{\citenamefont{Lidz et~al.}(2009)\citenamefont{Lidz, Zahn, Furlanetto,
  McQuinn, Hernquist, and Zaldarriaga}}]{Lidz:2008ry}
\bibinfo{author}{\bibfnamefont{A.}~\bibnamefont{Lidz}},
  \bibinfo{author}{\bibfnamefont{O.}~\bibnamefont{Zahn}},
  \bibinfo{author}{\bibfnamefont{S.}~\bibnamefont{Furlanetto}},
  \bibinfo{author}{\bibfnamefont{M.}~\bibnamefont{McQuinn}},
  \bibinfo{author}{\bibfnamefont{L.}~\bibnamefont{Hernquist}},
  \bibnamefont{and}
  \bibinfo{author}{\bibfnamefont{M.}~\bibnamefont{Zaldarriaga}},
  \bibinfo{journal}{Astrophys. J.} \textbf{\bibinfo{volume}{690}},
  \bibinfo{pages}{252} (\bibinfo{year}{2009}), \eprint{0806.1055}.

\bibitem[{\citenamefont{{Lidz} et~al.}(2011)\citenamefont{{Lidz}, {Furlanetto},
  {Oh}, {Aguirre}, {Chang}, {Dor{\'e}}, and {Pritchard}}}]{Lidz11}
\bibinfo{author}{\bibfnamefont{A.}~\bibnamefont{{Lidz}}},
  \bibinfo{author}{\bibfnamefont{S.~R.} \bibnamefont{{Furlanetto}}},
  \bibinfo{author}{\bibfnamefont{S.~P.} \bibnamefont{{Oh}}},
  \bibinfo{author}{\bibfnamefont{J.}~\bibnamefont{{Aguirre}}},
  \bibinfo{author}{\bibfnamefont{T.-C.} \bibnamefont{{Chang}}},
  \bibinfo{author}{\bibfnamefont{O.}~\bibnamefont{{Dor{\'e}}}},
  \bibnamefont{and} \bibinfo{author}{\bibfnamefont{J.~R.}
  \bibnamefont{{Pritchard}}}, \bibinfo{journal}{\apj}
  \textbf{\bibinfo{volume}{741}}, \bibinfo{eid}{70} (\bibinfo{year}{2011}),
  \eprint{1104.4800}.

\bibitem[{\citenamefont{{Gong} et~al.}(2012)\citenamefont{{Gong}, {Cooray},
  {Silva}, {Santos}, {Bock}, {Bradford}, and {Zemcov}}}]{Gong_2012}
\bibinfo{author}{\bibfnamefont{Y.}~\bibnamefont{{Gong}}},
  \bibinfo{author}{\bibfnamefont{A.}~\bibnamefont{{Cooray}}},
  \bibinfo{author}{\bibfnamefont{M.}~\bibnamefont{{Silva}}},
  \bibinfo{author}{\bibfnamefont{M.~G.} \bibnamefont{{Santos}}},
  \bibinfo{author}{\bibfnamefont{J.}~\bibnamefont{{Bock}}},
  \bibinfo{author}{\bibfnamefont{C.~M.} \bibnamefont{{Bradford}}},
  \bibnamefont{and} \bibinfo{author}{\bibfnamefont{M.}~\bibnamefont{{Zemcov}}},
  \bibinfo{journal}{\apj} \textbf{\bibinfo{volume}{745}}, \bibinfo{eid}{49}
  (\bibinfo{year}{2012}), \eprint{1107.3553}.

\bibitem[{\citenamefont{{Chang} et~al.}(2015)\citenamefont{{Chang}, {Gong},
  {Santos}, {Silva}, {Aguirre}, {Dor{\'e}}, and {Pritchard}}}]{Chang_2015}
\bibinfo{author}{\bibfnamefont{T.~C.} \bibnamefont{{Chang}}},
  \bibinfo{author}{\bibfnamefont{Y.}~\bibnamefont{{Gong}}},
  \bibinfo{author}{\bibfnamefont{M.}~\bibnamefont{{Santos}}},
  \bibinfo{author}{\bibfnamefont{M.~B.} \bibnamefont{{Silva}}},
  \bibinfo{author}{\bibfnamefont{J.}~\bibnamefont{{Aguirre}}},
  \bibinfo{author}{\bibfnamefont{O.}~\bibnamefont{{Dor{\'e}}}},
  \bibnamefont{and}
  \bibinfo{author}{\bibfnamefont{J.}~\bibnamefont{{Pritchard}}},
  \bibinfo{journal}{Advancing Astrophysics with the Square Kilometre Array
  (AASKA14)} \bibinfo{eid}{4} (\bibinfo{year}{2015}), \eprint{1501.04654}.

\bibitem[{\citenamefont{{Dumitru} et~al.}(2018)\citenamefont{{Dumitru},
  {Kulkarni}, {Lagache}, and {Haehnelt}}}]{Dumitru_2018}
\bibinfo{author}{\bibfnamefont{S.}~\bibnamefont{{Dumitru}}},
  \bibinfo{author}{\bibfnamefont{G.}~\bibnamefont{{Kulkarni}}},
  \bibinfo{author}{\bibfnamefont{G.}~\bibnamefont{{Lagache}}},
  \bibnamefont{and} \bibinfo{author}{\bibfnamefont{M.~G.}
  \bibnamefont{{Haehnelt}}}, \bibinfo{journal}{arXiv e-prints}
  (\bibinfo{year}{2018}), \eprint{1802.04804}.

\bibitem[{\citenamefont{{Beane} et~al.}(2018)\citenamefont{{Beane},
  {Villaescusa-Navarro}, and {Lidz}}}]{2018arXiv181110609B}
\bibinfo{author}{\bibfnamefont{A.}~\bibnamefont{{Beane}}},
  \bibinfo{author}{\bibfnamefont{F.}~\bibnamefont{{Villaescusa-Navarro}}},
  \bibnamefont{and} \bibinfo{author}{\bibfnamefont{A.}~\bibnamefont{{Lidz}}},
  \bibinfo{journal}{arXiv e-prints}  (\bibinfo{year}{2018}),
  \eprint{1811.10609}.

\bibitem[{\citenamefont{{Beane} and {Lidz}}(2018)}]{BL_2018}
\bibinfo{author}{\bibfnamefont{A.}~\bibnamefont{{Beane}}} \bibnamefont{and}
  \bibinfo{author}{\bibfnamefont{A.}~\bibnamefont{{Lidz}}},
  \bibinfo{journal}{\apj} \textbf{\bibinfo{volume}{867}}, \bibinfo{eid}{26}
  (\bibinfo{year}{2018}), \eprint{1806.02796}.

\bibitem[{\citenamefont{{Fujimoto} et~al.}(2019)\citenamefont{{Fujimoto},
  {Ouchi}, {Ferrara}, {Pallottini}, {Ivison}, {Behrens}, and
  {Gallerani}}}]{Fujimoto_2019}
\bibinfo{author}{\bibfnamefont{S.}~\bibnamefont{{Fujimoto}}},
  \bibinfo{author}{\bibfnamefont{M.}~\bibnamefont{{Ouchi}}},
  \bibinfo{author}{\bibfnamefont{A.}~\bibnamefont{{Ferrara}}},
  \bibinfo{author}{\bibfnamefont{A.}~\bibnamefont{{Pallottini}}},
  \bibinfo{author}{\bibfnamefont{R.~J.} \bibnamefont{{Ivison}}},
  \bibinfo{author}{\bibfnamefont{C.}~\bibnamefont{{Behrens}}},
  \bibnamefont{and}
  \bibinfo{author}{\bibfnamefont{S.}~\bibnamefont{{Gallerani}}},
  \bibinfo{journal}{arXiv e-prints}  (\bibinfo{year}{2019}),
  \eprint{1902.06760}.

\bibitem[{\citenamefont{{Bowman} et~al.}(2018)\citenamefont{{Bowman}, {Rogers},
  {Monsalve}, {Mozdzen}, and {Mahesh}}}]{Bowman_2018}
\bibinfo{author}{\bibfnamefont{J.~D.} \bibnamefont{{Bowman}}},
  \bibinfo{author}{\bibfnamefont{A.~E.~E.} \bibnamefont{{Rogers}}},
  \bibinfo{author}{\bibfnamefont{R.~A.} \bibnamefont{{Monsalve}}},
  \bibinfo{author}{\bibfnamefont{T.~J.} \bibnamefont{{Mozdzen}}},
  \bibnamefont{and} \bibinfo{author}{\bibfnamefont{N.}~\bibnamefont{{Mahesh}}},
  \bibinfo{journal}{\nat} \textbf{\bibinfo{volume}{555}}, \bibinfo{pages}{67}
  (\bibinfo{year}{2018}), \eprint{1810.05912}.

\bibitem[{\citenamefont{Chen and Miralda-Escude}(2004)}]{Chen:2003gc}
\bibinfo{author}{\bibfnamefont{X.-L.} \bibnamefont{Chen}} \bibnamefont{and}
  \bibinfo{author}{\bibfnamefont{J.}~\bibnamefont{Miralda-Escude}},
  \bibinfo{journal}{Astrophys. J.} \textbf{\bibinfo{volume}{602}},
  \bibinfo{pages}{1} (\bibinfo{year}{2004}), \eprint{astro-ph/0303395}.

\bibitem[{\citenamefont{Ewall-Wice et~al.}(2018)\citenamefont{Ewall-Wice,
  Chang, Lazio, Dore, Seiffert, and Monsalve}}]{Ewall-Wice:2018bzf}
\bibinfo{author}{\bibfnamefont{A.}~\bibnamefont{Ewall-Wice}},
  \bibinfo{author}{\bibfnamefont{T.~C.} \bibnamefont{Chang}},
  \bibinfo{author}{\bibfnamefont{J.}~\bibnamefont{Lazio}},
  \bibinfo{author}{\bibfnamefont{O.}~\bibnamefont{Dore}},
  \bibinfo{author}{\bibfnamefont{M.}~\bibnamefont{Seiffert}}, \bibnamefont{and}
  \bibinfo{author}{\bibfnamefont{R.~A.} \bibnamefont{Monsalve}}
  (\bibinfo{year}{2018}), \eprint{1803.01815}.

\bibitem[{\citenamefont{{Visbal} et~al.}(2015)\citenamefont{{Visbal}, {Haiman},
  and {Bryan}}}]{Visbal15}
\bibinfo{author}{\bibfnamefont{E.}~\bibnamefont{{Visbal}}},
  \bibinfo{author}{\bibfnamefont{Z.}~\bibnamefont{{Haiman}}}, \bibnamefont{and}
  \bibinfo{author}{\bibfnamefont{G.~L.} \bibnamefont{{Bryan}}},
  \bibinfo{journal}{\mnras} \textbf{\bibinfo{volume}{450}},
  \bibinfo{pages}{2506} (\bibinfo{year}{2015}), \eprint{1501.03177}.

\bibitem[{\citenamefont{{Gong} et~al.}(2013)\citenamefont{{Gong}, {Cooray}, and
  {Santos}}}]{Gong13}
\bibinfo{author}{\bibfnamefont{Y.}~\bibnamefont{{Gong}}},
  \bibinfo{author}{\bibfnamefont{A.}~\bibnamefont{{Cooray}}}, \bibnamefont{and}
  \bibinfo{author}{\bibfnamefont{M.~G.} \bibnamefont{{Santos}}},
  \bibinfo{journal}{\apj} \textbf{\bibinfo{volume}{768}}, \bibinfo{eid}{130}
  (\bibinfo{year}{2013}), \eprint{1212.2964}.

\end{thebibliography}

\end{document}